\documentclass[aps,prd,twocolumn,preprintnumbers,superscriptaddress,amsmath,floatfix,amssymb,showkeys]{revtex4}

\usepackage{multirow, graphicx,amssymb,url,mathrsfs,amsmath,amsfonts}
\usepackage{eucal,wrapfig,setspace}
\usepackage{amsxtra,amstext,latexsym}
\usepackage{pdflscape}
\usepackage{slashed}
\usepackage{graphicx}
\usepackage{xcolor}

\usepackage{caption}
\usepackage{sidecap}
\usepackage{dutchcal}
\usepackage{calligra}
 \newcommand{\scrn}{\mbox{${\mathscr N}$}}

\def\IR{{\hbox{{\rm I}\kern-.2em\hbox{\rm R}}}}
\def\IB{{\hbox{{\rm I}\kern-.2em\hbox{\rm B}}}}
\def\IN{{\hbox{{\rm I}\kern-.2em\hbox{\rm N}}}}
\def\IC{\,\,{\hbox{{\rm I}\kern-.59em\hbox{\bf C}}}}
\def\IZ{{\hbox{{\rm Z}\kern-.4em\hbox{\rm Z}}}}
\def\IP{{\hbox{{\rm I}\kern-.2em\hbox{\rm P}}}}
\def\IH{{\hbox{{\rm I}\kern-.4em\hbox{\rm H}}}}
\def\ID{{\hbox{{\rm I}\kern-.2em\hbox{\rm D}}}}

\newcommand{\beq}{\begin{equation}}
\newcommand{\eeq}{\end{equation}}
\newcommand{\bea}{\begin{eqnarray}}
\newcommand{\eea}{\end{eqnarray}}

\renewcommand{\topfraction}{1.0}
\renewcommand{\bottomfraction}{1.0}
\renewcommand{\textfraction}{0.0}
\begin{document}
%%%%%%%%%%%%%%%%%%%%%%%
\renewcommand{\topfraction}{1.0}
\renewcommand{\bottomfraction}{1.0}
\renewcommand{\textfraction}{0.0}
%%%%%%%%%%%%%%%%%%%%%%%

\newcommand\sect[1]{\emph{#1}---}

\title{Any Room Left for Technicolor? Holographic Studies of NJL Assisted Technicolour}

\author{Alexander Belyaev}
\email{a.belyaev@soton.ac.uk}
\affiliation{STAG Research Centre and  Physics \& Astronomy, University of
Southampton, Southampton, SO17 1BJ, UK}
\affiliation{Particle Physics Department, Rutherford Appleton Laboratory, Chilton, Didcot, Oxon OX11 0QX, UK}

\author{Kazem  Bitaghsir  Fadafan}
\email{bitaghsir@shahroodut.ac.ir}
\affiliation{Faculty of Physics, Shahrood University of Technology, P.O.Box 3619995161 Shahrood, Iran}

\author{Nick Evans}
\email{evans@soton.ac.uk} 
\affiliation{STAG Research Centre and  Physics \& Astronomy, University of
Southampton, Southampton, SO17 1BJ, UK}

\author{Mansoureh Gholamzadeh}
\email{mgh.gholamzadeh@shahroodut.ac.ir}
\affiliation{Faculty of Physics, Shahrood University of Technology, P.O.Box 3619995161 Shahrood, Iran}


\begin{abstract}
{We use a holographic description of technicolor dynamics to study gauge theories that only break chiral symmetry when aided by a strong four fermion interaction. These Nambu-Jona-Lasinio (NJL) assisted technicolor models provide examples of different dynamics from walking technicolor which can, by tuning, generate a light higgs like $\sigma$ meson. We compute the vector meson ($\rho$) and axial vector meson (A) spectrum for a variety of models with techni-quarks in the fundamental representation, enlarging the available parameter space over a previous analysis of walking theories. These predictions determine the parameter space of a low energy effective description  where LHC constraints from dilepton channels have already been applied. Many of the models with low numbers of electroweak doublets still lie beyond current constraints and motivate exploration of new signatures beyond dilepton for LHC and a 100 TeV proton collider.   }
\end{abstract}

\maketitle

\section{Introduction}

The discovery of a light higgs boson at 125 GeV \cite{Chatrchyan:2012xdj} and then the failure to find any new physics up to 1 TeV or above suggests nature is fine tuned. Models of Beyond the Standard Model physics are then hard to celebrate since they too are all uncomfortably tuned. Nevertheless it seems sensible to continue to attempt to rule out possible paradigms. The paradigm we will consider here is technicolor models \cite{Weinberg:1975gm} of electroweak symmetry breaking  which solve the hierarchy problem to the Planck scale by naturally generating a strong coupling regime in the TeV energy range and a resulting composite higgs. Any such electroweak dynamics must be rather different from QCD though since it must generate a light $\sigma$ particle to play the role of the higgs \cite{Appelquist:1998xf} and generate lower contributions to the precision S parameter \cite{Peskin:1990zt}. Examples of ideas that might change the dynamics are walking \cite{Holdom:1981rm}, where the anomalous dimension of the quark bilinear $\bar{q} q$, $\gamma$, runs slowly in the strong coupling regime, or the addition of strongly coupled Nambu-Jona-Lasinio four fermion operators \cite{Nambu:1961tp}. It has been hard to understand these new strongly coupled models since lattice work is intensive and struggles with widely separated scales (although much promising work has been done to move in this direction \cite{DeGrand:2015zxa}). 

Recently models have emerged from holography \cite{Maldacena:1997re} that potentially allow study of these strongly coupled systems. Holography provides a rigorous method of computation in a selection of strongly coupled gauge theories close to {$\scrn=4$ supersymmetric} gauge theory including theories with quarks \cite{Karch:2002sh}. If one works in the limit where one can neglect quark loops, the quenched (probe) limit, the key ingredient to determine the spectrum is precisely the running of $\gamma$ \cite{Jarvinen:2011qe}. Embracing that observation we can construct holographic models of generic gauge theories \cite{Alho:2013dka}. The predictions for the QCD ($N_c=3,N_f=2$) spectrum lie surprisingly close to observation at roughly the 10$\%$ level and one can hope as one moves away to theories with e.g. walking behaviour that the models will continue to make sensible predictions of the spectrum \cite{Erdmenger:2014fxa}. 

In a previous paper \cite{Belyaev:2018jse} we studied the spectrum and couplings of the $\rho$ and $A$ vector mesons of gauge theories with walking dynamics using our holographic model (light pseudo-Goldstone modes could also exist but when they do they are hard to pin down because their mass is determined by the potentially unknown origin of flavour physics and explicit singlet quark masses).  The philosophy here is to lean over backwards to construct a realistic model in order to understand whether it is, or how it could be, excluded. For this reason we picked a holographic model that supports the idea of a light $\sigma$ emerging if $\gamma$ runs slowly at the chiral symmetry breaking scale.  We then manhandled the IR running (that is unknown at strong coupling) to a precisely tuned form to generate the physical $\sigma$/higgs mass relative to the pion decay constant $F_\pi$ which determines the weak scale in these models. We also tuned a further parameter (the 5d gauge coupling $\kappa$ below) to set $S=0.1$ although this is only a tuning at one part in ten so not as extreme as that needed to get the higgs mass. 

The predictions for the spectrum of this holographic dynamics naturally fit very well in the parameter space of a low energy effective theory \cite{Foadi:2007ue} already widely studied \cite{Belyaev:2008yj}. That model has the Standard Model fields, higgs, and massive vector and axial gauge fields to play the role of the $\rho/V$ and $A$. The parameter space reduces, once the precision S parameter, higgs mass, $m_h$, and electroweak scale $v$ are fixed, to the $M_A$ versus $\tilde{g} = \sqrt{2} M_V/F_V$ plane (here $M_V, F_V$ are the mass and decay constant of the vector). There has been considerable study of limits on this plane from LHC dilepton final state data \cite{Belyaev:2008yj} - the $V$ and $A$ states mix with the Standard Model gauge bosons so can be singly produced and give very clean Drell Yan signals. The constraints had been reported as restrictive \cite{Belyaev:2018qye}, including exclusions over 3 TeV and for $\tilde{g}$ upto 8 at low masses. What our holographic analysis \cite{Belyaev:2018jse} showed though was, that at least for the dynamics in the holographic model, the top down predicted spectrum lay at large $\tilde{g} \sim 8.5$ and large $M_A \sim 4$ TeV. There is therefore still work to be done to exclude the paradigm because these parts of parameter space are relevant to top down models. Additional channels at LHC should be considered to add to the constraints and higher energy colliders may be needed to complete the exclusion (we are working on extending these limits and will report in a future publication).

{In both the holographic models and low energy effective description there is a third parameter 
\begin{equation}   \omega = {1 \over 2} \left({F_\pi^2 + F_A^2 \over F_V^2 } - 1  \right). \end{equation}
From the point of view of the LHC constraints we are interested in here, this parameter is unimportant if $|\omega|<0.3$ so we did not seek to tune it in the holographic model (where it typical lies around 0.05). However, it is worth stressing that in the low energy effective description which includes mixing with the electroweak gauge bosons this is tightly constrained to lie at the 0.001 level by the electroweak measurements of $\sin^2 \theta_w$. To make a holographic model fully compliant with constraints one needs an addition dial to tune - for example one could add higher dimension operators (by changing the UV boundary conditions in any given channel as we will see below) to represent instanton effects or high scale dynamics. Functionally though one is now talking about tuning in a three parameter space of the running strength, $\kappa$ and this new parameter at around 1 part in 1000 or more. This tuning {\it is} present already in the low energy effective description analysis and one can test there that changing $|\omega|$ up to 0.1 only changes the spectrum at the few percent level. The goal of our holographic analysis is to ask a top down model to broadly indicate the predicted pieces of parameter space of that model one would expect from a UV completion of the physics. Here fine tunings which effect the spectrum at a few percent level are not crucial (the tuning of the light higgs mass is the most important tuning and we do include that) and the conclusion that the high $M_A$ and large $\tilde{g}$ volume of parameter space are important certainly remains. The goal of highlighting the need for probes of that volume remains also. }

A natural question to ask is how robust are the predictions of the holographic model?  This is hard to answer since it is a model and not rigorously derived from the base theory described. {Indeed it is worth stressing that even though a holographic model may capture QCD at the 15\% level if treated favourably by fitting across all parameters the errors can as much as double if one fits to a single parameter. The holographic models are a very coarse predictor. We are simply using them here to motivate new pieces of parameter space to encourage experimental searches.}

To add to the understanding of the variance in the predictions of the holographic model, here we want to ask how the predictions are effected if the underlying dynamics is changed. Recent work has shown how the dynamics of the Nambu-Jona-Lasinio model can be easily incorporated into the holographic description \cite{Jarvinen:2015ofa,Evans:2016yas,BitaghsirFadafan:2018efw}. Thus the models we will study here will consist of a technicolour theory that has not reached the critical coupling needed for chiral symmetry breaking at the electroweak scale. The symmetry breaking nevertheless occurs because of the aid of a four fermion interaction term at a scale of a few 10s of TeV with a tuned coupling. This dynamics is easily incorporated into the holographic description using Witten's multi-trace prescription \cite{Witten:2001ua} and the expected dynamics is observed. In a previous paper we also showed that if the running of the gauge theory is sufficiently slow then a light higgs can result in these models \cite{BitaghsirFadafan:2018efw}. Such models have been previously called NJL assisted Technicolour (NJL-aTC) or Ideal Walking models \cite{Fukano:2010yv, Lane:2016kvg}.  

Here we will analyse models with an SU($N_c$) gauge theory and techni-quarks in the fundamental representation. We first concentrate on models with a single electroweak doublet present (which formally the experimental constraints we will use apply to) but with varying number of singlet techni-quarks to vary the total number of flavours $N_f$, and hence the running in the theory. The techni-flavours $\psi^a$ also have common four fermion/ NJL interactions
\beq {\cal L}_{NJL} = {g^2 \over \Lambda^2} \left| \bar{\psi}^a_L \psi^a_R \right|^2 \label{njllag}\eeq
with $g^2$ the coupling and $\Lambda$ the UV cut off, which act to enhance the interaction for techni-quark condensation. We explain how to include these interactions holographically in Section 2. The holographic model again contains a coupling ($\kappa$) that can be tuned to generate the benchmark value $S=0.1$.

To generate the light higgs in the holographic model we need the running of the gauge coupling or anomalous dimension of the $\bar{\psi} \psi$ condensate to be very carefully tuned (at one part in 100) to a sufficiently gentle running value at the symmetry breaking scale. There are potentially two such solutions (shown in one case in Figure 1 below). One is present for any gauge theory where it is somewhat weakly coupled as it leaves the asymptoticaly free regime. The second possibility is if in the strongly coupled regime it approaches an IR fixed point at a value of $\gamma$ that is sub-critical - in this case there is a strongly coupled solution. We explore both possibilities, identifying models tuned to the physical $F_\pi, m_h$ and S and computing the properties of the $\rho,A$ mesons {(we again neglect the extra level of tuning to set $\omega$ small to keep the numerics tractable)}. We concentrate on models with a UV cut of 20 times the technicolor scale as a simple, plausible example (higher cut offs don't show very different behaviours). {For the weakly coupled models the values of $M_A$ and $\tilde{g}$ lie at $M_A$ lower than those we saw in the walking theories - they are more typical of a scaled up QCD theory ($M_A \sim 3.25$ TeV, $\tilde{g} \sim 7$). The model predicts high values of $\tilde{g} \simeq 10$ though.} On the other hand, for the most strongly coupled theories the $M_A$ value is moved higher to as much as 4 TeV. Our interpretation here is that the strongly coupled, slowly walking theories connect the low scale dynamics to the higher cut off scale of the NJL dynamics more strongly and this tends to raise the A mass. These strongly coupled NJL-assisted theories appear in the same region of the $M_A$ parameter space as the strongly coupled walking theories of our previous paper, supporting this understanding. The theories with weaker gauge coupling but stronger NJL coupling are therefore the key point of this paper - they provide alternative UV dynamics that leaves the theory in different, smaller $M_A$ regions of the low energy effective theory's parameter space. By varying $N_f$ we can move smoothly between these regimes. This analysis therefore enlarges the region of parameter space in the low energy theory that true UV completions can realize. Unfortunately, for the one doublet models, this area of parameter space still lies beyond the reach of LHC dilepton limits.  {From the models we construct here it is clear that essentially the entire region where $M_V> M_A$ and $M_A<4$TeV would need to be excluded experimentally.}

It is also worth stressing that the walking models of \cite{Belyaev:2018jse} may very well not exist if the tuned IR runnings needed to generate the light higgs are not the true dynamics. On the other hand these NJL assisted models have sub-critical gauge dynamics with running periods that surely exist and are  consistent and realizable theories at least below the scale of the NJL interactions $\Lambda$.

We also study the predictions of similarly tuned models with more than one electroweak doublet - here the experimental limits are expected to be similar to those on one doublet models although they have not been explicitly determined.  Increasing the number of doublets enlarges the electroweak $F_\pi$ relative to the rest of the spectrum and therefore, at fixed weak scale, tends to decrease both $M_A$ and less so $\tilde{g}$ bringing the theories closer to exclusion. 

These realizations of the UV technicolor theories we  have  found,  are  mostly beyond  the  reach
of the current LHC dilepton searches.  However, they do
motivate new signatures (which we will explore in future work) but also the reach of future colliders
with higher energies. Here we present projections for dilepton searches at  
14 TeV (3 ab$^{-1}$), 27 TeV (15 ab$^{-1}$)  and 100 TeV (3 ab$^{-1}$)  $pp$ colliders which begin to probe the parameter space of the models.

\section{Holographic Model}
\subsection{Holographic Description of Gauge Dynamics}

We use a holographic model to describe the gauge dynamics - the Dynamic AdS/QCD model which is described in detail in \cite{Alho:2013dka} and originates from the D3/probe D7 system. The action is
\begin{equation} \begin{array}{lcl}
S & = & -\int d^4x~ d u {\rm{Tr}}\, u^3 
\left[  {1 \over r^2} |D X|^2 \right.  \\ &&\\
&& \left. \hspace{1cm}  +  {\Delta m^2(r) \over u^2} |X|^2   + {1 \over 2 \kappa^2} (F_V^2 + F_A^2) \right], 
\label{daq} \end{array}
\end{equation}
Here $u$ is the holographic coordinate dual to energy scale.

$X$ is the field dual to the quark condensate $\bar{q} q$. $\Delta m^2$ is a renormalization group scale/radially dependent mass term which we fix using the two loop running of the gauge coupling in the theory of interest as described in \cite{Alho:2013dka} - this ansatz, $\Delta m^2 =  -2 \gamma = -3 \alpha(u)(N_c^2-1)/2 N_c \pi$ includes IR fixed points for the running for appropriate choices of $N_c,N_f$. Chiral symmetry breaking is induced if $\Delta m^2$  falls through -1 where there is an instability (the Brietenlohmer Freedman bound \cite{Breitenlohner:1982jf} is violated).

The solution of $X$'s equation of motion, which can be found numerically for a given choice of $\Delta m^2$, describes the vacuum of the theory. We use the on mass shell condition $|X|(u=X_0) = X_0$ with $|X|'(X_0)=0$ and require $|X|=m$ in the UV or at the cut off $\Lambda$ to describe a techni-quark of bare mass $m$. Asymptotically the solution is $|X| = m + \langle \bar{q}_L q_R \rangle / u^2$ so the bare mass and quark condensate can be extracted numerically. Fluctuations of $X$ describe the $\sigma$ (or higgs) and $\pi$ fields. 

The vector and axial vector fields describe the operators $\bar{q} \gamma^\mu q$ and $\bar{q} \gamma^\mu \gamma_5 q$ and their fluctuations give the $\rho$ and {$A$} spectrum and couplings.  The five dimensional coupling $\kappa$ is a free parameter of the holographic description.

The theory lives in a geometry
\begin{equation} ds^2 = r^2 dx_{3+1}^2 + {1 \over r^2} du^2, ~~~~~~~r^2 = u^2 + |Tr X|^2 \end{equation} 
 $|Tr X|$ is included in the definition of $r$ in the metric which provides a ``back-reaction'' on the metric in the spirit of probe brane models \cite{Karch:2002sh} and communicates the mass gap to the mesonic spectrum.

The spectrum of the theory is found by looking at linearized fluctuations of the fields about the vacuum where fields generically take the 
form $f(u)e^{ip.x}, p^2=-M^2$. The resulting Sturm-Louville equation for $f(u)$ gives a discrete spectrum. By substituting the wave functions back into the action and integrating over $u$ the decay constants can also be determined.
The normalizations of the fluctuations are determined by matching to the gauge theory expectations for the vector-vector, axial-axial and scalar-scalar correlators in the UV of the theory as described in detail in \cite{Alho:2013dka}. Note that in the holographic literature \cite{{Alho:2013dka}} the dimension 2 coupling between the vector meson and it's associated source is normally written as $F_V^2$ whilst in the Weinberg sum rule literature \cite{Peskin:1990zt} it is written as $m_V F_V$. We will adopt the latter definition here to fit the other literature on technicolor.

Our models will focus first on a single electroweak doublet of techni-quarks but we will assume the existence of technicolor singlet quarks to change the UV running of the coupling. In the computations of $F_\pi$ and $F_{V/A}$ for the electroweak physics only the electroweak doublet contributes - so factors of $N_f$ and $N_c$ in these quantities reflect the values in a one doublet model.  We will also present results for models with 2,3 and 4 doublets.

We will  tune our models to the observed electroweak data so their predictions for the $V,A$ spectrum can be excluded. In this spirit we will tune $\kappa$ to produce {$\rho$--$A$} degeneracy to ensure the electroweak $S$ parameter \cite{Peskin:1990zt}
\begin{equation}
S=4\pi\left[\frac{F_V^2}{M_V^2}-\frac{F_A^2}{M_A^2}\right] \ , \label{eq:WSR0} 
\end{equation}
is sufficiently small  (we pick $S=0.1$ as a benchmark point). Note tuning $\kappa$ to zero makes the Lagrangian terms for the $\rho$ and $A$ the same so that the $A$ mass drops to that of the $\rho$. However, since the suppressed, first term in the action is the one which links the symmetry breaking $X$ to the $A$, to maintain $F_\pi^2$ (which is the leading value in the $AA$ correlator) one must raise the overall scale.

\subsection{Holographic NJL Operators}

Our main goal in this paper is to include NJL four techni-quark interactions into the holographic model so that they contribute to the chiral symmetry breaking dynamics. Does this change the predictions for the $V,A$ masses relative to the pure, walking gauge dynamics we described in \cite{Belyaev:2018jse}?

NJL interactions can be introduced simply using Witten's double trace prescription \cite{Witten:2001ua} - see \cite{Evans:2016yas} for a top down analysis of these operators in the D3/probe D7 system and for formal studies of their impact on gauge dynamics.

 The basic principle is that if we add a Lagrangian  term to the gauge theory of the form in (\ref{njllag})
 and then condensation occurs so $\langle \bar{\psi}_L \psi_R \rangle \neq 0$ then an effective hard mass is generated at the UV scale $\Lambda$
 \begin{equation} m_\Lambda =  {g^2 \over \Lambda^2} \langle \bar{\psi}_L\psi_R \rangle \label{getg} \end{equation}
 Witten's prescription in our model is to allow solutions for $X(u)$ with non-zero mass, $m$, at the scale $\Lambda$ but interpret them as the massless theory with the NJL term present. For these solutions one can read off the associated $g^2$ of the NJL interaction by substituting the numerical values of $m$ and $\langle \bar{\psi}_L\psi_R \rangle$ from $X$ into (\ref{getg}).

\subsection{$N_c=3, N_f=12$ Example}
 
Let us consider a particular example of this dynamics at work. We will pick a theory with $N_c=3$ and $N_f=12$ (that is one electroweak doublet and 10 electroweak singlet quarks) and describe it with the Dynamic AdS/QCD model based on the two loop running of the gauge coupling. The running of $\gamma$ is shown in Figure 1 -  note this theory lies in the conformal window and is never quite strong enough in this ansatz to break chiral symmetry by itself (that is $\Delta m^2$ in the holographic model never falls to -1).  The theory can be made to break chiral symmetry though by the inclusion of an NJL term at a scale $\Lambda$.
 
We pick two scales - one is $X_0$ the IR value of the holographic field $X$ which can be thought of as the IR mass scale induced by the symmetry breaking. We also pick a UV cut off $\Lambda$ for example at 20 $X_0$. 
We now link these scales to the gauge dynamics by picking $\gamma_{IR}=\gamma(X_0)$ - this can be any choice a priori so that the theory samples any running profile over a factor of 20 in scales from the range of running of the full gauge dynamics. In Figure 1 we show two examples of such ranges by the vertical lines. We solve for $X(\rho)$ between $X_0$ and $\Lambda$.
 
We can next compute the spectrum and couplings of this background as usual. Here one must pick a value for $\kappa$ - for the moment one can pick any value. Note one needs to be careful when looking at excitations of $X$ that represent the $\sigma$ or higgs - asymptotically the $u$ dependent wave functions for these  fluctuations $\delta X(u)$ must satisfy the same ratio of mass to condensate (ie NJL coupling $g^2$) at $\Lambda$ as the background field $X(u)$. This condition can be enacted numerically as the UV boundary condition rather than the usual $\delta X(\Lambda)=0$. Now one can make this an electroweak symmetry breaking model by imposing $F_\pi=246$GeV and rescaling all other scales appropriately.  Generically of course the $\sigma$ mass will not be 125 GeV. 

The next step is to vary $\gamma_{IR}$ to attempt to find a theory where the $\sigma$ mass does also match the observed higgs mass. That is, one attempts to find a period of running that is suitably tuned flat to generate the correct higgs mass.  In this case there are two such regimes which we will describe shortly.
 
Even having completed this program of computations one will find that the S parameter is not our chosen value of 0.1. Now one can return through this cycle but varying $\kappa$ until both $m_h$ and $S$ are the desired values - this is a laborious task!  

{As mentioned in the Introduction, at this point one could also embark to tune $\omega$ to zero. One could for example introduce higher dimension operators associated with $\rho$ or $A$ operators by changing the UV boundary conditions in the meson computations. One could justify this as including instanton effects in the strongly coupled gauge theory because the large $N_c$ origin of holography neglects these effects. Or given the presence of an NJL operator here one might expect extra higher dimension operators from the same origin. Tuning at 1 part in 1000 in a 3 dimensional space though is too numerically intensive for the benefit. This final tuning is not expected to change the spectrum sufficiently (based on studying changes in the spectrum when $|\omega|$ is changed by up to 0.1 in the low energy theory) to change our conclusions. }

In Table 1 we show the outcome of the computations after tuning $m_h$ and S - for the moment consider just the two entries for $N_f=12$. The two regions of running between $X_0$ and $20X_0$ are shown in Figure 1 - the regions between the two vertical lines. The key thing to achieve the correct higgs mass in the model is that the running is ``just right'' being neither too steep nor too gentle. There is therefore one solution to each side of the steepest running period in the model. 

\begin{figure}[htb]
   \includegraphics[width=0.4\textwidth]{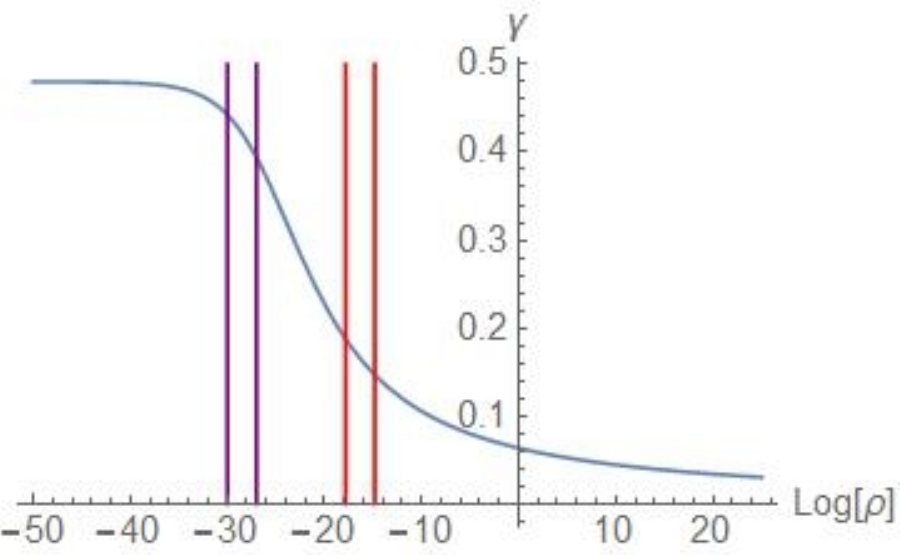}
  \caption{\label{fig:1}  Here we show the running of the anomalous dimension, $\gamma$, against RG scale $\rho$, for the $N_C=3$ $N_f=12$ theory. It asymptotes to an IR value that places it, using the 2 loop $\beta$ function, just inside the conformal window before chiral symmetry breaking sets in. Here we also show the IR and UV scales, separated by a factor of 20, of two NJL assisted theories that generate $m_h/F_\pi = 0.5$.
  } \end{figure}

We can now look at the behaviour of the model as $N_f$ is changed. The two loop runnings as a function of $N_f$ and $N_c$ provide ansatzes for the non-perturbative running of $\gamma$. Of course they are not trustable since they are based on perturbation theory but they are indicative of behaviours we expect to see. Given this we will choose to use fractional $N_f$ to explore the possible behaviours as a continuous parameter is changed in the running (at large $N_c$ $N_f/N_c$ does become continuous). Now again consider Table 1.  

\begin{table}[!htbp]
		\begin{tabular}{cc|cc|cccc|c|cc|cc}
			\hline $N_c$ & $N_f$ & $\kappa$ & $\gamma_{IR}$  & $M_A$ GeV & $\tilde{g}$  & $\omega$ \\ \hline
			3 & 2. & 1.88 & 0.087 & 3173.84 & 9.48985 & 0.021 \\
			3 & 7. & 2.02 & 0.106 & 3155.96 & 9.55277 & 0.026 \\
			3 & 8. & 2.06 & 0.113 & 3157.07 & 9.57466 & 0.027 \\
			3 & 9. & 2.12 & 0.121 & 3151.88 & 9.60021 & 0.029 \\
			3 & 10. & 2.21 & 0.133 & 3145.25 & 9.63716 & 0.032 \\
			3 & 10.6 & 2.27 & 0.142 & 3146.26 & 9.66491 & 0.034 \\
			3 & 11. & 2.33 & 0.15 & 3143.9 & 9.68823 & 0.036 \\
			3 & 11.2 & 2.36 & 0.1548 & 3146.69 & 9.70334 & 0.036 \\
			3 & 12 & 2.61 & 0.188 & 3148.39 & 9.7985 & 0.042 \\
			3 & 12.2 & 2.75 & 0.204 & 3146.36 & 9.84262 & 0.045 \\
			3 & 12.4 & 3. & 0.236 & 3162.9 & 9.92767 & 0.049 \\
			3 & 12.4 & 3.61 & 0.314 & 3239.04 & 10.1171 & 0.057 \\
			3 & 12.2 & 4.21 & 0.383 & 3323.12 & 10.2644 & 0.061 \\
			3 & 12 & 4.75 & 0.441 & 3405.94 & 10.3749 & 0.063 \\
			3 & 11.2 & 7.55 & 0.696 & 3844.55 & 10.7299 & 0.070 \\
			3 & 11. & 8.6 & 0.7758 & 3989.28 & 10.8029 & 0.074 \\
			3 & 10.6 & 11.7 & 0.9695 & 4322.82 & 10.9215 & 0.091 \\
			\hline
		\end{tabular}
		\caption{Data from the holographic model for $N_c=3$ with one doublet $(N_D=1)$ and $\Lambda_{UV}/m_{IR}=20$, and with $\gamma_{IR}$ and $\kappa$ tuned to $m_h = f_\pi/2$ and S=0.1
		\label{tab:nc3_nd1_l20}}
\end{table}

Let us first consider increasing $N_f$ above 12. The IR fixed point value falls as $N_f$ increases and the running generically becomes slower. The two solutions which give the correct $m_h$ value move inwards towards the point of strongest running - that is the $\gamma_{IR}$ values for the two solutions move towards each other. Before one reaches $N_f$=13 the two solutions merge at the point of strongest running and beyond that the running becomes too slow to generate a large enough $m_h$.

If one reduces $N_f$ the IR fixed point rises, the running becomes stronger in the middle regime,  and soon the gauge theory is capable of breaking chiral symmetry unaided. Before this happens we see the two NJL assisted solutions separate (the $\gamma_{IR}$s move apart) as they move away from the strongest running  regime (which generates too large an $m_h$). Further the solution at strongest coupling quite soon ends as $N_f$ decreases because the running at the scale where the symmetry breaking occurs (where $\gamma_{IR}$ is now very close to 1) is too strong and the higgs too heavy.  
The weaker coupling solution continues to exist for all $N_f$ but is pushed into the UV to ensure that the running remains sufficiently slow. By $N_f$=2 the $\gamma_{IR}$ value has fallen below 0.1 which is only very barely non-perturbative and holographic modelling could no longer be sensible.  For these ``weaker coupling solutions'' the NJL dynamics is dominating the symmetry breaking    whilst for the ``stronger coupling solutions'' the gauge theory is playing a much larger role. 

Note there is also the possibility of ``NJL opposed'' symmetry breaking  where a repulsive (negative $g^2$) NJL interaction acts against the gauge dynamics - see \cite{BitaghsirFadafan:2018efw} for a discussion of this mechanism in  gauge theories and 
\cite{BitaghsirFadafan:2018iqr} for discussions in a condensed matter context.
Mostly where we have found such solutions,  they have too large a higgs mass (in the holographic model to achieve a negative $g^2$ the solution $X(u)$ must turn negative before the cut off - such solutions are very $u$ dependent and this breaking of conformal symmetry leads to a large $m_h$). The most strongly coupled points (at largest $M_A$) for our theories are briefly of this type so we include such solutions. 

In summary, for $N_c$=3 we have understood how to holographically describe NJL assisted technicolor models and found weaker (at all $N_f$) and stronger coupling solutions (near $N_f=12$) that generate the correct $m_h$ and $S$. Explicit numerical results are provided for this case in Table~\ref{tab:nc3_nd1_l20} and the values are plotted in the $M_A-\tilde{g}$ plane in Figure 2 (this plane will be the phenomenologically interesting plane below). Note the U-shaped nature of the curve with varying (continuous) $N_f$ - the cusp corresponds to where the two solutions described above merge at the largest value of $N_f=12.4$. The branch that moves to highest $M_A$ is the strongly coupled solutions at smaller $N_f$ up to the point where they cease to exist at $N_f=10.6$. The lower $M_A$ branch are weakly coupled solutions which exist for all $N_f$ below $12.4$. 

One can also study the dependence of the theory on the UV cut off value. For example, consider increasing $\Lambda$ by an order of magnitude from 20 to 200 times the IR scale for the case of $N_c=3, N_f=12$ (for $\Lambda=200 X_0$ numerical data is provided in the Table~\ref{tab:nc3_nd1_l200} of the appendix). We show the equivalent results to those discussed above in Figure 2. Note that the results still have a ``two branch meeting at a cusp'' structure but here the two branches lie on top of each other in the plot.  We find the $\rho$ and $A$ meson masses for the weakly coupled solutions tend to move lower by as much as 500 GeV although the effect is less pronounced on the stronger coupled solutions (which are strongly linked to the higher cut off scale). We have also considered higher cut offs but the results are largely unchanged from the $\Lambda/X_0 = 200$ case. {These very low values of $M_A$ appear inconsistent with the low energy theory below - in the holographic model $M_A>M_V$ whilst in the low energy theory $M_V>M_A$ at these values of $M_A$. The origin of this discrepancy is that $\omega$ has grown quite large in the holographic models' results and here one should undertake the extra tuning skipped above to reduce it. Given, as we will see, that the models at $\Lambda=20 X_0$ already motivate searches across the full allowed $M_A$ range with $M_V>M_A$ we again do not pursue this tuning. }
Henceforth we will present results just for $\Lambda/X_0 = 20$ since they present a robust challenge to collider constraints but one should remember that changing the cut off can influence the precise predictions (but not our generic conclusions below). 

\begin{figure}[htb]
   \includegraphics[width=0.5\textwidth]{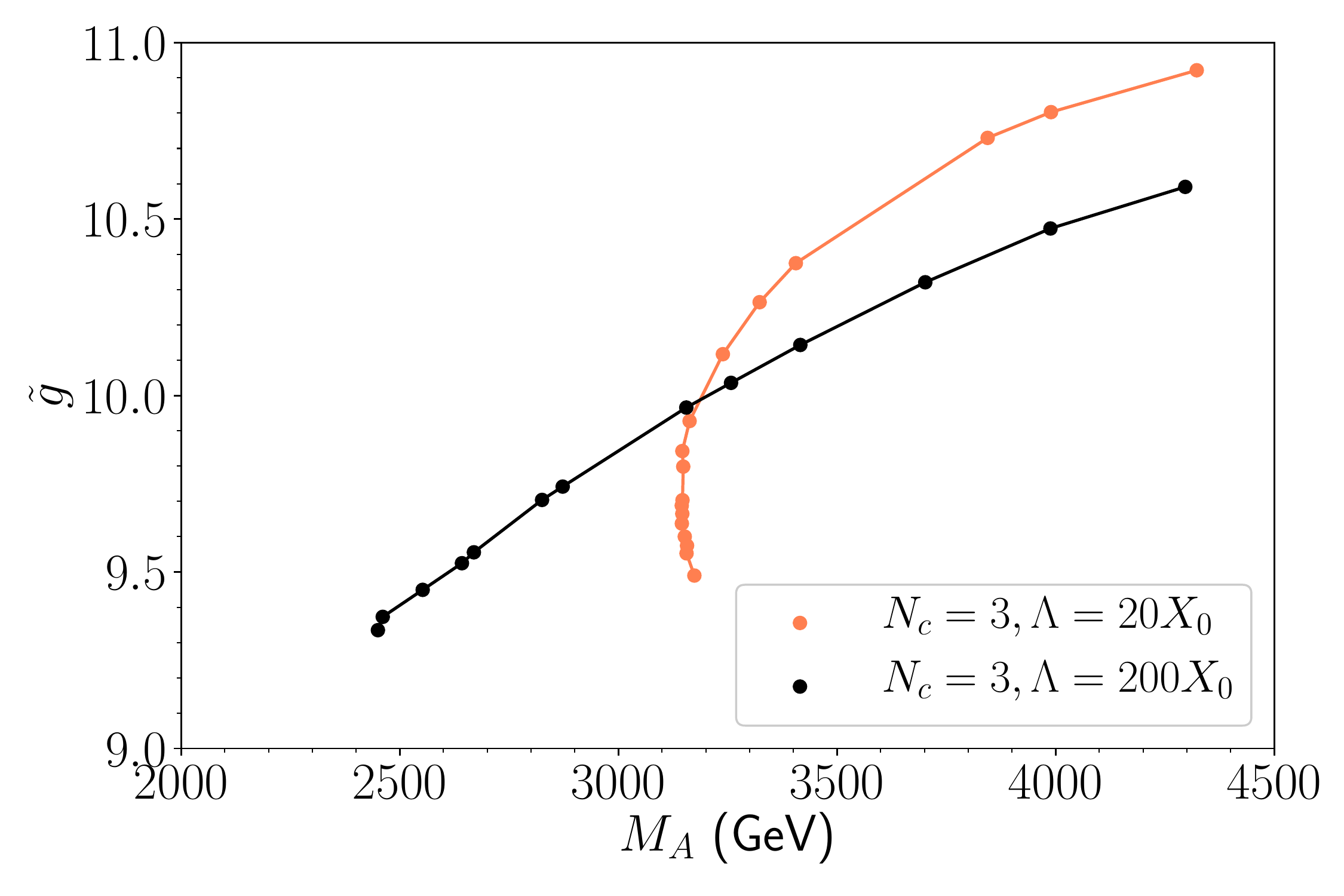}
  \caption{\label{fig:2}  The solutions from Table~\ref{tab:nc3_nd1_l20} for the $N_c=3$ theory in the $M_A$ vs $\tilde{g}$ plane. The orange points are for $\Lambda=20 X_0$, the black ones are for $\Lambda=200 X_0$ .
  } \end{figure}

For the phenomenological analysis below we will use this holographic technology to explore the predictions of a range of models model with different choices of  $N_c$ and $N_f$.

 \subsection{Parameter Counting}
 
 The parameter count in the holographic model is worth noting: for a particular theory with $N_c,N_f$ the  running of $\alpha$ (and hence the anomalous dimension $\gamma$) is fixed by the perturbative two loop result.
 We choose the ratio $X_0/\Lambda$ which effectively determines $F_\pi/\Lambda$. We then set the scale of the theory by requiring $F_\Pi=246$~GeV   and fix $\gamma_{IR}$   through $m_h$. 
 The model then predicts $M_\rho, F_\rho, M_A, F_A$. We next tune $\kappa$ to force S=0.1. The remaining three predictions we will express as
\begin{equation} M_A,  ~~~~~ \tilde{g} = {\sqrt{2} M_V \over F_V},   ~~~~  \omega = {1 \over 2} \left({F_\pi^2 + F_A^2 \over F_V^2 } - 1  \right). \end{equation}
In fact for all our models $\omega < 0.15$ which is at a level where the experimental constraints are unchanged in  the high energy reach regime so we suppress that parameter in our plots.

\section{Phenomenological Model} 
We will make use of the phenomenological model of the spin {one} states made from a single electroweak doublet from \cite{Foadi:2007ue} as we did in
\cite{Belyaev:2018jse}. It is important to stress that the same low energy theory describes the walking theories we considered previously and the NJL assisted models we study here. Although the dynamics of chiral symmetry breaking is changing the low energy degrees of freedom and symmetries are the same. The Lagrangian for the low energy effective theory is 

\begin{eqnarray}
{\cal L}_{\rm boson}&=&-\frac{1}{2}{\rm Tr}\left[\widetilde{W}_{\mu\nu}\widetilde{W}^{\mu\nu}\right]
-\frac{1}{4}\widetilde{B}_{\mu\nu}\widetilde{B}^{\mu\nu}\nonumber \\ &
-&\frac{1}{2}{\rm Tr}\left[F_{{\rm L}\mu\nu} F_{\rm L}^{\mu\nu}+F_{{\rm R}\mu\nu} F_{\rm R}^{\mu\nu}\right] \nonumber \\
&+& m^2\ {\rm Tr}\left[C_{{\rm L}\mu}^2+C_{{\rm R}\mu}^2\right]
+\frac{1}{2}{\rm Tr}\left[D_\mu M D^\mu M^\dagger\right] \nonumber \\ &
-& \tilde{g}^2\ r_2\ {\rm Tr}\left[C_{{\rm L}\mu} M C_{\rm R}^\mu M^\dagger\right] \nonumber \\
&-&\frac{i\ \tilde{g}\ r_3}{4}{\rm Tr}\left[C_{{\rm L}\mu}\left(M D^\mu M^\dagger-D^\mu M M^\dagger\right) \right.\nonumber \\ 
&+&  \left. C_{{\rm R}\mu}\left(M^\dagger D^\mu M-D^\mu M^\dagger M\right) \right] \nonumber \\
&+&\frac{\tilde{g}^2 s}{4} {\rm Tr}\left[C_{{\rm L}\mu}^2+C_{{\rm R}\mu}^2\right] {\rm Tr}\left[M M^\dagger\right]\nonumber \\ &
+& \frac{\mu^2}{2} {\rm Tr}\left[M M^\dagger\right]-\frac{\lambda}{4}{\rm Tr}\left[M M^\dagger\right]^2
\label{eq:boson}
\end{eqnarray}
where $\widetilde{W}_{\mu\nu}$ and $\widetilde{B}_{\mu\nu}$ are the ordinary electroweak field gauge fields, $F_{{\rm L/R}\mu\nu}$ are the field strength tensors associated to the vector meson fields $A_{\rm L/R\mu}$~\footnote{In Ref.~\cite{Foadi:2007ue}, where the chiral symmetry is SU(4) there is an additional term whose coefficient is labelled $r_1$. With an SU($N$)$\times$SU($N$) chiral symmetry this term is just identical to the $s$ term.}, and the $C_{{\rm L}\mu}$ and $C_{{\rm R}\mu}$ gauge fields are
$C_{{\rm L}\mu}\equiv A_{{\rm L}\mu}-\frac{g}{\tilde{g}}\widetilde{W_\mu}$ and \\ $
C_{{\rm R}\mu}\equiv A_{{\rm R}\mu}-\frac{g^\prime}{\tilde{g}}\widetilde{B_\mu}
$

The matrix $M$  takes the form
\begin{eqnarray}
M=\frac{1}{\sqrt{2}}\left[v+H+2\ i\ \pi^a\ \tau^a\right]\ ,\quad\quad  a=1,2,3
\end{eqnarray}
Here $\pi^a$ are the Goldstone bosons produced in the chiral symmetry breaking, $v=\mu/\sqrt{\lambda}$ is the corresponding VEV, and $H$ is the composite higgs. The higgs is assumed to have Standard Model Yukawa couplings to the fermions. The covariant derivative is
\begin{eqnarray}
D_\mu M&=&\partial_\mu M -i\ g\ \widetilde{W}_\mu^a\ \tau^a M + i\ g^\prime \ M\ \widetilde{B}_\mu\ \tau^3\ . 
\end{eqnarray}
When $M$ acquires its VEV, the Lagrangian of Eq.~(\ref{eq:boson}) contains mixing matrices for the spin one fields. The mass eigenstates are the ordinary SM bosons, and two triplets of heavy mesons:$\rho$ and  $A$.

Including all the interactions with the electroweak gauge and higgs fields of dimension 4 needs six parameters: the mass, $m$ and coupling $\tilde{g}$ of the new gauge fields, the higgs VEV $v$, and three couplings $r_2, r_3$ and $s$. 
The model predicts 
\begin{equation}
M_V^2 = m^2 + \frac{\tilde{g}^2\ (s-r_2)\ v^2}{4}, ~~~~~
M_A^2 = m^2 + \frac{\tilde{g}^2\ (s+r_2)\ v^2}{4} 
\label{eq:masses}
\end{equation}
and
\begin{equation}
F_V  =  \frac{\sqrt{2}M_V}{\tilde{g}} \ , 
F_A  =  \frac{\sqrt{2}M_A}{\tilde{g}}\chi \ ,
F_\pi^2  =  \left(1+2\omega\right)F_V^2-F_A^2 \ ,
\label{eq:FVFAFP}
\end{equation}
where
\begin{eqnarray}
\omega \equiv \frac{v^2 \tilde{g}^2}{4 M_V^2}(1+r_2-r_3) \ , \quad \quad 
\chi \equiv 1-\frac{v^2\ \tilde{g}^2\ r_3}{4 M_A^2} \ . \label{eq:chi}
\end{eqnarray}
{Without loss of generality we chose $s=0$ here, noting that:
a) the $\rho/A$ 
production rates, as well as the partial decay width of
Z to fermions (di-jets and dileptons) are independent of $s$ (at the per-mil level);
b)  the branchings of $\rho$ to} {dileptons} {increases by 10$\%$ at most for
$s$ reaching 10 in absolute value because of the  $\rho \to ZH$  partial width decreases;
c) we do not involve here higgs boson phenomenology and use only the dilepton channel to probe the phenomenological model parameter space.
%Setting $s=0$ therefore still provides robust constraints on the theory.
}

Of the five remaining variables we set $F_\Pi=$246~GeV, and $S=$0.1. This leaves three degrees of freedom $M_A, \tilde{g}, \omega$ which can be experimentally constrained.

Phenomenologically the three parameters are treated as completely free parameters. 
The parameter count is the same as that of the holographic model which makes absolute predictions for these numbers as a function of $N_c, N_f$ and $\Lambda$.

In \cite{Belyaev:2008yj} the model was implemented in  CalcHEP~\cite{Belyaev:2012qa} using LanHEP~\cite{Semenov:2014rea} to automatically derive the Feynman rules.
This  implementation is publicly available at HEPMDB (High Energy Physics Model Database) ~\cite{hepmdb} under hepmdb:1012.0102 ID.
In \cite{Belyaev:2018jse} the model
implementation was extended  by nonzero $s$ and $\omega$ parameters and became available under hepmdb:1218.0319. In our study we use this implementation of the model.

The most relevant signals from  the model 
come from $\rho/A$ resonant production either in Drell-Yan(DY) or Vector Boson Fusion (VBF) production processes followed by  $\rho/A$
decay to fermions or boson. Here we use
$pp\to \rho/A \to \ell^+\ell^-$ DY process with dilepton (di-electron and di-muon) final states as the easiest (but not necessarily the best, as we discuss below) one for reinterpretation of the present experimental data and projections for future luminosities and collider energies.
We use a very similar  approach as in \cite{Belyaev:2018qye}.

We have explored the dependence of the experimental constraints on the parameter $\omega$. For $|\omega| < 0.3$ the impact on the exclusion regime is small and any changes occur at $M_A \simeq 1.5$~TeV. The high mass reach area is least {affected}. Given the holographic models, both for the walking theories in \cite{Belyaev:2018jse} and our NJL assisted models here place $\omega < 0.15$ in all cases we will simply suppress this parameter which is not playing a significant role in constraining the models.

A further useful parameter to monitor (although it is not independent) is  $\mathcal{a}$ from
\begin{equation} \mathcal{a} 4 \pi^2 F_\pi^4 = F_\rho^2 M_\rho^2 - F_A^2 M_A^2 \end{equation}
which allows tracking   of the second Weinberg sum rule or equally the degeneracy of the $\rho-A$ pair.

\section{Results}

Our first goal is to explore the  LHC's potential to probe the NJL-aTC model space, as predicted by the holographic model,  using recent limits  from the dilepton DY search channel. 
We use LHC  searches for  dilepton resonances only
and reinterpret them for the NJL-aTC parameter space. The choice of dilepton signature is very well motivated since this is  the cleanest signature to search for  the vector resonances. However, as we will see below (and as it was shown in~\cite{Belyaev:2018jse}) it becomes less efficient in the region of large values of $\tilde g$ where couplings of  resonances to SM fermions are suppressed.

\begin{figure}[htb]
   \includegraphics[width=0.5\textwidth]{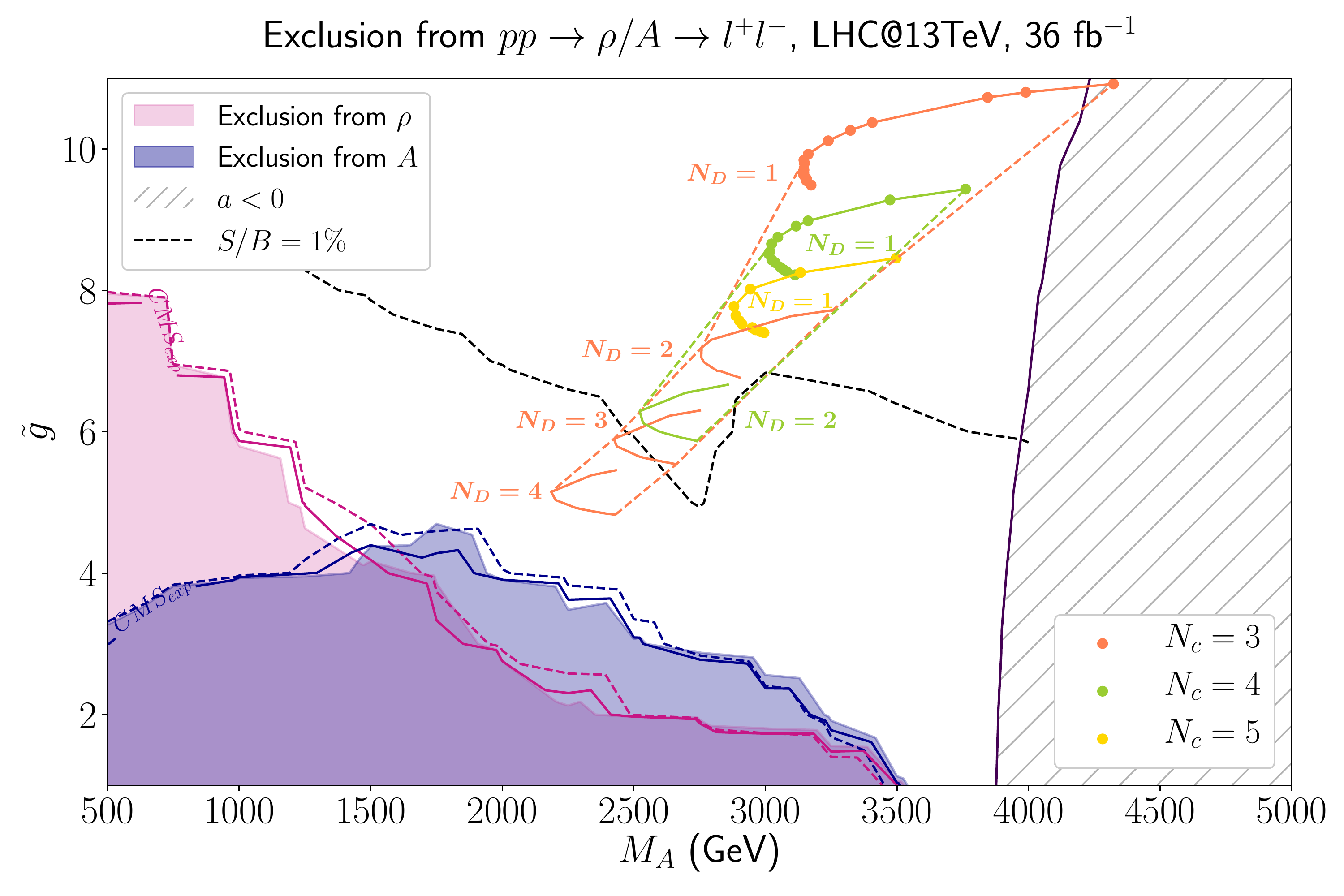}
  \caption{\label{fig:collider}
 The combination of shaded (pink and dark blue) regions presents 95\% CL 
exclusion  of the $M_A-\tilde{g}$ space 
 from the CMS observed  limit on  dilepton resonance searches at the LHC@13TeV with 36 fb$^{-1}$. Solid and dashed lines along the borders of the shaded area represent an expected CMS limit and our 
 limit  using binned likelihood method respectively.
 The predictions for NJL-ATC  holographic model with a cut off of 20 times the IR techniquark mass(tuned at each $N_c,N_f,$ to 
give $S$=0.1 and the correct higgs mass) are overlaid.  The  coral-coloured 
dots are  predictions for $N_c=3$ and different $N_f$, while green and gold are 
for  $N_c=4$ and $N_c=5$ cases respectively.
For $N_c=3$ we present ``trajectories" for 
higher number of the EW doublets: $N_D=2,3,4$
and connect them by the dashed line to indicate overall region of theory space.
Parameter $\mathcal{a}$, from the phenomenological model, is related to 
$\rho-A$ degeneracy and the holographic points lie near the line $\mathcal{a}=0$ 
as a result of tuning to a small $S$ parameter. 
  } \end{figure}

In Figure~\ref{fig:collider} we present the up-to-date LHC reach for the model parameter space.
We use here the CMS { DY limits on $Z^{\prime}$ production at 13~TeV ($36 fb^{-1}$) from the dilepton (combined dielectron and dimuon) final state~\cite{Sirunyan:2018exx} reinterpreted as limits on the phenomenological model's  parameter space. The CMS limit in \cite{Sirunyan:2018exx} was expressed 
as a ratio, $R_{\sigma}$ = $\sigma(pp\rightarrow Z^{\prime}\rightarrow 
\ell^{+}\ell^{-})/\sigma(pp\rightarrow Z\rightarrow \ell^{+}\ell^{-})$ -  the
$Z^{\prime}$ signal cross section in the dilepton final state to the Standard 
Model (SM) cross section of a Z boson to the dilepton final state. The choice of 
this ratio  $R_{\sigma}$ was made to remove the dependency on the theoretical 
prediction of $\sigma(pp\rightarrow Z\rightarrow \ell^{+}\ell^{-})$ and 
correlated experimental uncertainties.} {We have reinterpreted this limit (see  
details below)  and  have found  the phenomenological model's  $\tilde g-M_A$ parameter space 
excluded at 95\% confidence level(CL). We have found exclusion separately for  
$\rho$ and $A$ resonances, indicated by pink and dark-blue shaded regions 
respectively in Figure~\ref{fig:collider}. The  overlay of these regions gives 
us eventually the  overall exclusion region.
We have used CalcHEP to evaluate the signal dilepton signal cross sections at tree-level and the modified ZWPROD program~\cite{Accomando:2010fz} to evaluate the mass-dependent QCD NNLO  K-factor. } 

To make  projections for future collider energies and luminosities, first,  we have reproduced the CMS expected limits
using  the following binned likelihood method.
We assume resonance widths are negligible compared to the gaussian-smearing effects of finite detector resolution. The signal hypothesis probability density function is defined by a Gaussian distribution with the width equal to the detector resolution (1.2\% of resonance mass), and a signal-strength modifier, $\mu$, which is the expected number of events at the experiment. The background is estimated by  invariant dilepton mass distribution
with very high statistics. Where there are few background events (e.g. $m_{\ell^{+}\ell^{-}}\geq$ 2~TeV at 13~TeV), we use the CL$_s$ method alongside a toy Mont\'e Carlo in order to construct the distribution of a single test-statistic for background only and signal+background hypotheses.
One can see that the limit from our approach which is represented by dashed  lines along the borders of the shaded area in Figure~\ref{fig:collider} closely reproduces  the expected CMS limit 
from the dilepton search. This agreement  validates our approach and allows us  to use it  for future collider energies and luminosities projections which we present below.

In Figure~\ref{fig:collider} we also present a dashed black line lying in the large $\tilde g$ region and indicating a 1\% level of signal-to-background (S/B) ratio. This line gives an idea (from the most optimistic expected control of systematic uncertainties)
about the potential limit of  the dilepton signature  to probe the phenomenological model parameter space.

This contour line is not expected to  change with the increase of the collider energy since the irreducible dilepton background and the signal will scale the same way with the energy increase.
The dip for  $M_A\simeq 2.7$ TeV and $\tilde{g}\simeq 5$ in S/B contour is related to the sharp increase of $Br(\rho\to W^+W^-)$, which respectively leads to a sharp decrease of dilepton signal rate. This sharp change happen for  $M_A\simeq2.7$, where 
$\rho$ and $A$ are switching their roles:
in  $M_A> 2.7$ region $\rho$ becomes mostly vector, while $A$ becomes mostly pseudo-vector. It is important to note that the details here most likely depend on the fact that the analysis of the phenomenological theory has been done for one electroweak doublet of technifermions. For higher numbers of doublets there are more electroweak charged $\rho/A$s involving  respective mixing. The LHC  potential to probe the respective parameter space is likely to be enhanced in the case of   higher $\rho/A$   multiplicty, thus one can consider the limit in the $M_A$ ${\tilde g}$ parameter space for one doublet we are using here as a conservative one. We concentrate on one doublet models here which are hardest to experimentally exclude but it would be interesting to perform proper   analysis of the phenomenological model beyond one doublet models in the future.

Figure~\ref{fig:collider}  shows that the LHC limit reaches $M_A\simeq  3.5$~TeV (for small $\tilde{g}$) 
and $\tilde{g} \sim 8$ (for small $M_A$), so appears to be  tightly constraining. However, to orient ourselves in theory space consider a technicolor model that is a scaled up version of QCD, a theory whose spectrum we know.  We scale $f_\pi=93$~MeV to $F_\Pi = 246$~GeV and find $M_\rho= 2.05$~TeV, $M_A= 3.25$ TeV, $S=0.3$ and $\tilde{g}=7$. This theory is excluded by $S$ and the absence of a light higgs but provides a reference values to place on the exclusion plot Figure~\ref{fig:collider}. It is not excluded purely in terms of the $\rho,A$ bounds.   

The goal of our previous paper \cite{Belyaev:2018jse} was to ask where true models of technicolor lie in the larger parameter space of the phenomenological model - the particular dynamics of a top down construction should lie at just one point in the parameter space. In that paper we used our  holographic description to explore if there  are walking gauge theories, with manhandled IR running couplings, that generate the observed higgs mass. We found candidate walking models lay near $a \simeq 0$ (the edge of the hashed region in Figure 3 of this paper). For one electroweak doublet models $\tilde{g}$  lay near 8.5 for $N_c=3$ and  6.5  for $N_c=5$. As more electroweak doublets were added $\tilde{g}$ fell to as low as 3 for 6 doublet models. Here these walking models are strongly coupled over scales far above the electroweak scale and this had the effect of increasing the $A$ mass relative to the QCD case.  See \cite{Belyaev:2018jse}  for more details.

{Now we can add to these predictions by including NJL-aTC models. We show the theories predictions as data points in Figure \ref{fig:collider}. We considered for these plots  models with a UV cut off 20 times the IR techni-quark mass. For a given value of $N_c$ and number of electroweak doublets there is a trajectory along which $N_f$ changes. For example, consider the top trajectory in Figure 2 - this corresponds to $N_c=3$ and one electroweak doublet.  The data is therefore that in Table~\ref{tab:nc3_nd1_l20} and Figure 2. As we saw in Section IIC there are two solutions at some $N_f$ values. The first branch are relatively weakly coupled solutions where the NJL interaction dominates the physics - these solutions are bunched up near $M_A= 3.25$ close to the scaled up QCD value in this plane. These solutions have a higher $\tilde{g}$ as large as 10 though.  The trajectory to larger $M_A$ represents the stronger coupled solutions for $10.6 < N_f < 12.4$ (with $N_f=10.6$ the most strongly coupled solution and at largest $M_A$) - here the gauge dynamics does most of the work of breaking electroweak symmetry with a small assist form the NJL interaction. These solutions, as they get stronger, track  to the $a=0$ line where the equivalent $N_c=3$ walking theory in \cite{Belyaev:2018jse} lay. Again these theories are strongly coupled all the way to the UV cut off so the dynamics raises the $A$ mass. This seems like a consistent picture suggesting that possible top down completions of technicolor with a light higgs lie between scaled up QCD and a roughly 4 TeV $A$ mass. Our models here have predicted a larger value of $\tilde{g} \simeq 10$ than previously studied models and this presents a further challenge to experimental constraints. }

In Figure 3 we show  further trajectories for one doublet theories with $N_c=4$ and 5 (the data is in the Tables~\ref{tab:nc4_nd1_l20} and \ref{tab:nc5_nd1_l20} of the appendix).
{The increase in $N_c$ naturally decreases the mass scale of the theory since more colours are contributing to $F_\pi$ which is the physical weak scale. The holographic model predicts that $\tilde{g}$ will also decrease a little. Nevertheless the strongly coupled branch sees an increase in $M_A$ towards the $a=0$ contour again.   

For completeness we also show equivalent trajectories for the $N_c=3$ theory with 2,3 and 4 electroweak doublets (the respective data is given in Tables~\ref{tab:nc3_nd2_l20}-\ref{tab:nc3_nd4_l20} of the appendix) and for the $N_c=4$ theory with 2 electroweak doublets (the respective data is given in Table~\ref{tab:nc4_nd2_l20} of the appendix). Formally the bounds from the phenomenological model do not apply since it is presented and has been constrained only for the one doublet case. However, the results show that increasing the number of doublets contributing to $F_\pi$ does lower the scale of the new physics and leaves these models more open to experimental probing. 

Finally of course we should comment that our results here further emphasise the conclusion of \cite{Belyaev:2018jse} that top down models of technicolor with a small number of doublets and dynamics tuned to give a light higgs are not yet constrained by LHC searches.

\section{Beyond LHC}

\begin{figure}[htb]
   \includegraphics[width=0.5\textwidth]{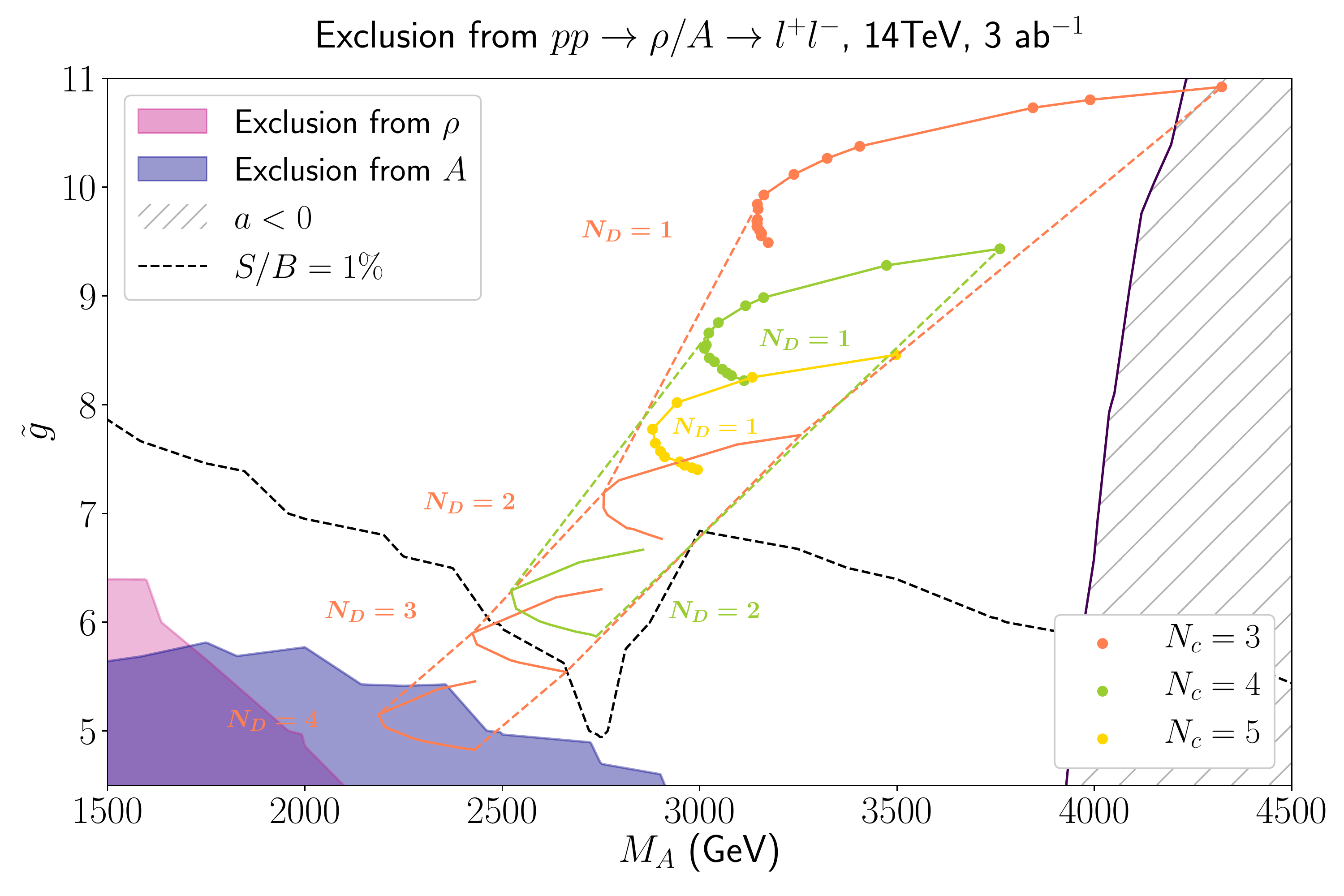}\\
    \includegraphics[width=0.5\textwidth]{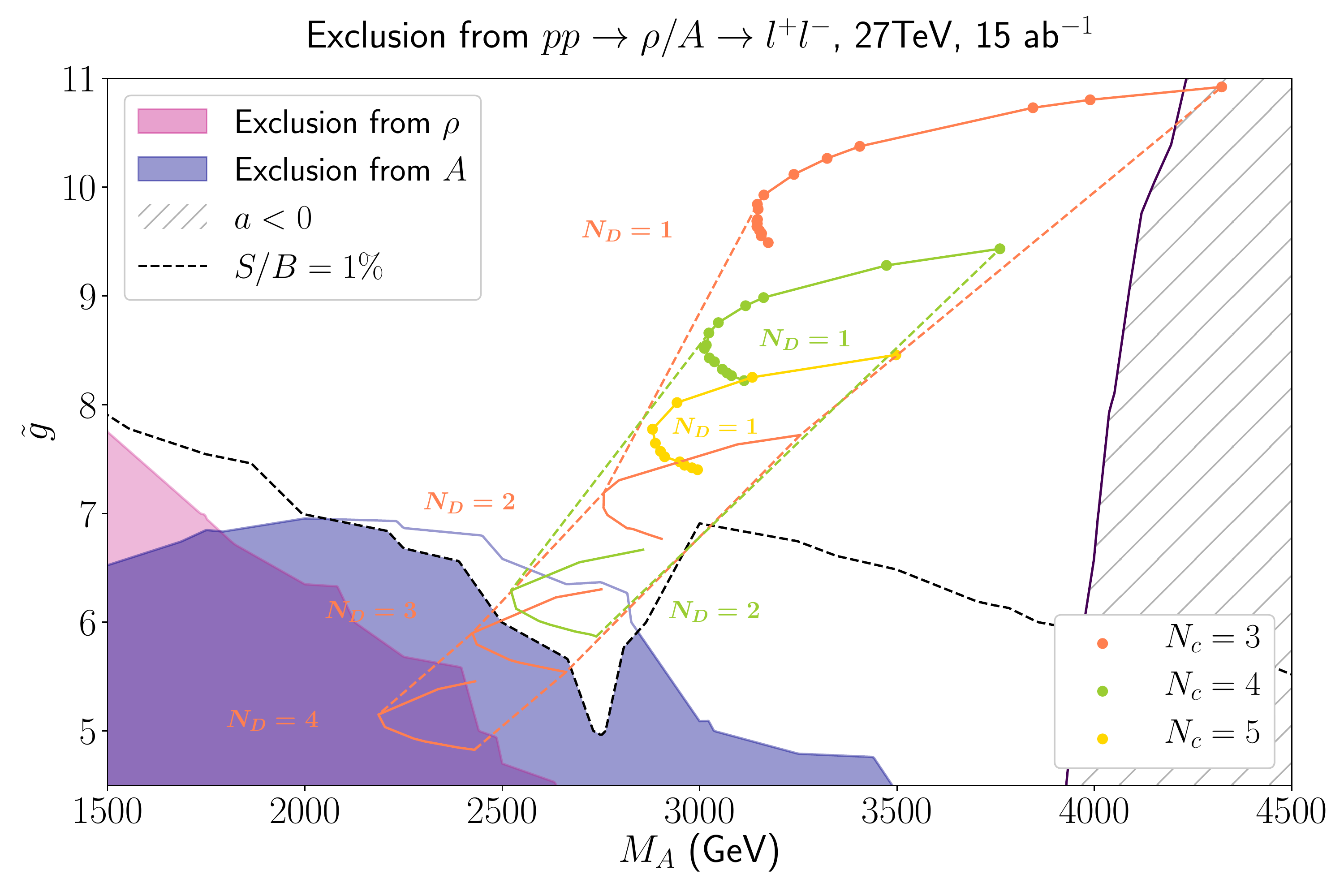}\\  
   \includegraphics[width=0.5\textwidth]{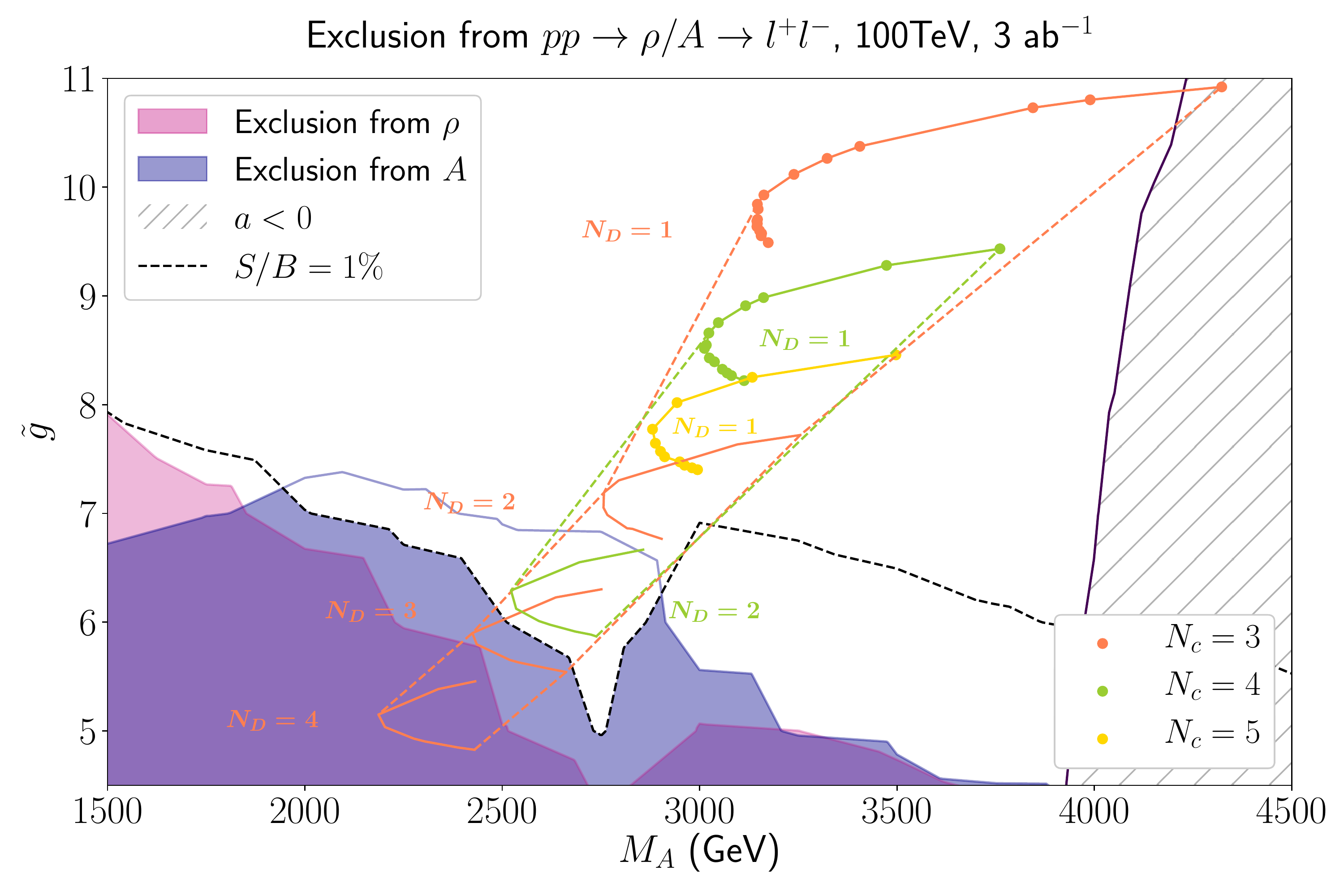}
  \caption{\label{fig:collider2}
  Shaded areas present 95\% CL projected exclusion  on the $M_A- \tilde{g}$ plane for 14(3 ab$^{-1}$)(top), 27(15 ab$^{-1}$) (middle) and 100 TeV (3 ab$^{-1}$)(bottom) $pp$ collider from  dilepton DY resonance searches. The notations are the same as in Figure~\ref{fig:collider}.
  } \end{figure}

We have found that the LHC dilepton searches  to date do not  exclude any region of parameter space of the walking or NJL assisted paradigms.  A total exclusion would need not only a higher collider energy but also new signatures to probe 3-4 TeV resonances especially in the large $\tilde{g}$ region
 with very low dilepton rates. We illustrate this point   in  Figure~\ref{fig:collider2} where we present projections for dilepton searches at  
14 TeV (3 ab$^{-1}$)(top), 27 TeV (15 ab$^{-1}$)(middle)  and 100 TeV (3 ab$^{-1}$)(bottom)  $pp$ colliders. One can see a gradual  improvement of the colliders potential while moving from the 14 TeV to 100 TeV case. One can see that the 14 TeV (3 ab$^{-1}$)
LHC will be able to probe the $N_D=4$ region of NJL-aTC parameter space, which in comparison to the holographic predictions for the walking theories~\cite{Belyaev:2018jse} for large $N_D$ are  shifted 
not only to the lower $\tilde{g}$ values but also to the lower $M_A$ values. 
Moreover, one can see that  27 TeV and 100 TeV $pp$ colliders will be able to probe the NJL-aTC parameter space for $N_c=3$ with $N_D \gtrsim 3$ and
 $N_c=4$ with $N_D\gtrsim 2$.

At the same time one can see that all of the models with $N_D=1$ lying in the $\tilde{g}>7$ and $M_A>3$~TeV region  are out of reach of even a 100~TeV collider if only the dilepton DY signature is used. 
 
 It is apparent that the dilepton signature becomes less efficient in probing the technicolor parameter space for large values of $\tilde g $ where the couplings of the $\rho/A$  to fermions are suppressed.
 Therefore exploration of higher values of $\tilde g $ motivates the study of additional di-boson signatures
 either from  DY production or from the additional VBF production channel.
 One should note that VBF production of $\rho/A$ followed by respective diboson(VV) or boson-higgs(VH) decay
 looks particularly promising in the very large  $\tilde g \simeq 7$ region since neither production nor decay of new heavy resonances are suppressed by $1/\tilde g$. 
 Moreover, the increase of collider energy can further enhance the significance of the VBF channel, which highlights a special role for the 100 TeV pp collider to potentially search the entire technicolor parameter space.

 \section{Discussion}
  
In this paper we have continued our analysis from \cite{Belyaev:2018jse} asking whether collider searches for techni-$\rho$ and A mesons can be used to exclude the technicolor paradigm. We are using recently developed holographic technology to efficiently compute the spectrum of a wide range of theories {(although we again caution that the errors on the predictions could be sizable - they serve to motivate collider searches though).}  The low energy description of such models with one electroweak doublet has been studied and there are constraints from dilepton Drell Yan processes at the LHC \cite{Belyaev:2008yj}. Figure 3 displays these constraints in this context. 

In \cite{Belyaev:2018jse} we studied theories where we allowed walking technicolor a last gasp chance to survive by adjusting by hand the IR running at the level of 1 part in 100  to generate a light $\sigma$ to play the role of the higgs. Such theories lived on the $a=0$ contour in Figure 3 and are not yet excluded. 

Here we have studied NJL assisted technicolor models where the gauge dynamics is sub-critical for chiral symmetry breaking but a four fermion interaction at a cut off $\Lambda$ (which we have set at roughly 20 TeV) helps drive electroweak symmetry breaking. A light higgs is again predicted if the gauge coupling's running is sufficiently slow but here this tuning can be achieved by choosing an appropriate period of the running between the weak scale and the cut off. This provides an alternative set of dynamics to test the robustness of the previous predictions. Theories of this type with strong gauge dynamics typically lie towards the $a=0$ boundary in Figure 3 as the walking theories did. On the other hand theories where the gauge coupling is much weaker and the NJL coupling drives the symmetry breaking dynamics lie at lower $M_A$ values that are somewhat more accessible, if not to LHC, then to future colliders (see Figure 4 for the improved bounds from dilepton DY).

{It is worth stressing that the  holographic models reveals the fine tuning needed (1 part in 1000 if one include tuning $m_H,S$ and $\omega$) to realize such a strongly coupled model of electroweak symmetry breaking. This is of course as unpalatable as in any other tuned model. Nevertheless it seems worthwhile to exclude the paradigm. We conclude that the full large $M_A$ parameter space upto of order 4 TeV in the low energy effective theory is of interest and attempts should be made to experimentally exclude it. }

In addition to motivating future colliders, these results again highlight a need to
go beyond dilepton DY channel and to
enlarge the analysis with additional (VBF)
production and di-boson/boson-higgs decays channels. Such advancing of the analysis will be the subject of our future publication and  could potentially lead to a discovery  or closure of the  technicolor paradigm.

\acknowledgements{
The authors are very grateful to
Daniel Locke for  discussions and help with establishing the LHC limits.
KBF would like to thank CERN for their hospitality and Johanna Erdmenger for hosting his visit to University of Wurzburg.
{AB's and NE's work was supported by the
STFC  consolidated  grant  ST/P000711/1}.
AB acknowledges support from Soton-FAPESP grant, ICTP-SAIFR Institute 
and Invisibles-plus programme. 
}

\section{Appendix}

 	\begin{table}[!htbp]
 		\begin{tabular}{cc|cc|cccc|c|cc|cc}
 			\hline $N_c$ & $N_f$ & $\kappa$ & $\gamma_{IR}$  & $M_A$ GeV & $\tilde{g}$ & $\omega$ \\ \hline 			
 		 3 & 7. & 1.07 & 0.113 & 2903.8 & 6.76636 & 0.029 \\
 		 3 & 9 & 1.15 & 0.131 & 2871.05 & 6.80515 & 0.034 \\
 		 3 & 10.6 & 1.27 & 0.156 & 2829.65 & 6.85821 & 0.041 \\
 		 3 & 11. & 1.32 & 0.1665 & 2816.17 & 6.86352 & 0.04 \\
 		 3 & 12. & 1.59 & 0.221 & 2766.36 & 6.98571 & 0.056 \\
 		 3 & 12.2 & 1.77 & 0.258 & 2756.48 & 7.05298 & 0.060 \\
 		 3 & 12.2 & 2.22 & 0.342 & 2757.74 & 7.19161 & 0.077 \\
 		 3 & 12. & 2.65 & 0.42 & 2795.2 & 7.30377 & 0.086 \\
 		 3 & 11. & 4.82 & 0.771 & 3094.87 & 7.63437 & 0.113 \\
 		 3 & 10.6 & 6.13 & 0.967 & 3255.41 & 7.72081 & 0.129 \\
 			\hline
 		\end{tabular}
 		\caption{Data from the holographic model for $N_c=3$ two doublets $(N_D=2)$ with $\Lambda_{UV}/m_{IR}=20$, and with $\gamma_{IR}$ and $\kappa$ tuned to $m_h = f_\pi/2$ and S=0.1
 		\label{tab:nc3_nd2_l20}}
 	\end{table}

	\begin{table}[!htbp]
		\begin{tabular}{cc|cc|cccc|c|cc|cc}
			\hline $N_c$ & $N_f$ & $\kappa$ & $\gamma_{IR}$  & $M_A$ GeV & $\tilde{g}$  & $\omega$ \\ \hline
		 3 & 7. & 0.84 & 0.126 & 2661.76 & 5.54162 & 0.035 \\
		 3 & 9. & 0.92 & 0.146 & 2613. & 5.5766 & 0.041 \\
		 3 & 10.6 & 1.06 & 0.178 & 2541.91 & 5.62969 & 0.050 \\
		 3 & 11. & 1.12 & 0.192 & 2519.49 & 5.65212 & 0.054 \\
		 3 & 12. & 1.55 & 0.289 & 2435.46 & 5.79584 & 0.075 \\
		 3 & 12. & 1.93 & 0.37 & 2424.33 & 5.90026 & 0.089 \\
		 3 & 11. & 3.96 & 0.765 & 2635.85 & 6.22799 & 0.129 \\
		 3 & 10.6 & 5.06 & 0.9638 & 2751.28 & 6.30215 & 0.144 \\							
			\hline
		\end{tabular}\caption{Data from the holographic model for $N_c=3$ three doublets $(N_D=3)$ with $\Lambda_{UV}/m_{IR}=20$, and with $\gamma_{IR}$ and $\kappa$ tuned to $m_h = f_\pi/2$ and S=0.1\label{tab:nc3_nd3_l20}}
	\end{table}

 	\begin{table}[!htbp]
 		\begin{tabular}{cc|cc|cccc|c|cc}
 			\hline $N_c$ & $N_f$ & $\kappa$ & $\gamma_{IR}$  & $M_A$ GeV & $\tilde{g}$  & $\omega$  \\ \hline
 		 3 & 8 & 0.78 & 0.148 & 2429.57 & 4.82693 & 0.043 \\
 		 3 & 9 & 0.84 & 0.163 & 2387.25 & 4.84825 & 0.047 \\
 		 3 & 10.6 & 1. & 0.2026 & 2304.46 & 4.90318 & 0.058 \\
 		 3 & 11. & 1.08 & 0.223 & 2275.79 & 4.92974 & 0.063 \\
 		 3 & 11.8 & 1.44 & 0.308 & 2202.82 & 5.03582 & 0.082 \\
 		 3 & 11.8 & 1.92 & 0.416 & 2186.3 & 5.15059 & 0.100 \\
 		 3 & 11. & 3.55 & 0.759 & 2337.84 & 5.38471 & 0.135 \\
 		 3 & 10.6 & 4.57 & 0.9607 & 2431.51 & 5.45628 & 0.152 \\				
 			\hline
 		\end{tabular}
 		\caption{Data from the holographic model for $N_c=3$ four doublets $(N_D=4)$ with $\Lambda_{UV}/m_{IR}=20$, and with $\gamma_{IR}$ and $\kappa$ tuned to $m_h = f_\pi/2$ and S=0.1
 		\label{tab:nc3_nd4_l20}}
 	\end{table}
 
 \clearpage

  	\begin{table}[!htbp]
  		\begin{tabular}{cc|cc|cccc|c|cc|cc}
  			\hline $N_c$ & $N_f$ & $\kappa$ & $\gamma_{IR}$  & $M_A$ GeV & $\tilde{g}$ & $\omega$  \\ \hline
  		
  		 4. & 2. & 1.38 & 0.088 & 3112.67 & 8.22107 & 0.020 \\
  		 4. & 8. & 1.48 & 0.1042 & 3080.76 & 8.26713 & 0.025 \\
  		 4. & 10. & 1.53 & 0.113 & 3069.71 & 8.29139 & 0.027 \\
  		 4. & 12. & 1.6 & 0.1256 & 3057.63 & 8.32556 & 0.031 \\
  		 4 & 14.3 & 1.75 & 0.152 & 3037.64 & 8.39431 & 0.037 \\
  		 4 & 15. & 1.84 & 0.166 & 3024.66 & 8.42963 & 0.040 \\
  		 4 & 16 & 2.06 & 0.202 & 3012.34 & 8.51722 & 0.047 \\
  		 4 & 16.2 & 2.13 & 0.215 & 3017.05 & 8.54779 & 0.050 \\
  		 4 & 16.5 & 2.45 & 0.265 & 3023.5 & 8.65904 & 0.057 \\
  		 4 & 16.5 & 2.75 & 0.311 & 3047.51 & 8.75399 & 0.063 \\
  		 4 & 16.2 & 3.33 & 0.396 & 3116.56 & 8.90973 & 0.071 \\
  		 4 & 16 & 3.67 & 0.442 & 3162.05 & 8.98434 & 0.075 \\
  		 4 & 15. & 5.6 & 0.684 & 3473.55 & 9.28038 & 0.088 \\
  		 4 & 14.3 & 7.7 & 0.909 & 3760.93 & 9.43319 & 0.102 \\
  		  					
  			\hline
  		\end{tabular} 
  		\caption{Data from the holographic model for $N_c=4$ one doublet $(N_D=1)$ with $\Lambda_{UV}/m_{IR}=20$, and with $\gamma_{IR}$ and $\kappa$ tuned to $m_h = f_\pi/2$ and S=0.1
  		\label{tab:nc4_nd1_l20}}
  	\end{table}

  	\begin{table}[!htbp]
  		\begin{tabular}{cc|cc|cccc|c|cc|cc}
  			\hline $N_c$ & $N_f$ & $\kappa$ & $\gamma_{IR}$  & $M_A$ GeV & $\tilde{g}$ & $\omega$ \\ \hline
  		4 & 8 & 0.89 & 0.121 & 2738.32 & 5.869 & 0.033 \\
  		4 & 10 & 0.92 & 0.129 & 2724.87 & 5.88569 & 0.036 \\
  		4 & 12 & 0.99 & 0.145 & 2684.93 & 5.91411 & 0.040 \\
  		4 & 14.3 & 1.14 & 0.1795 & 2621.76 & 5.97608 & 0.050 \\
  		4 & 15. & 1.23 & 0.1995 & 2593.19 & 6.01003 & 0.055 \\
  		4 & 16. & 1.56 & 0.2715 & 2535.26 & 6.12375 & 0.071 \\
  		4 & 16. & 2.18 & 0.395 & 2522.24 & 6.29143 & 0.090 \\
  		4 & 15. & 3.64 & 0.673 & 2695.61 & 6.55105 & 0.118 \\
  		4 & 14.3 & 4.95 & 0.904 & 2856.95 & 6.66686 & 0.136 \\
  			  		 						
  			\hline
  		\end{tabular}
  		\caption{Data from the holographic model for $N_c=4$ two doublets $(N_D=2)$ with $\Lambda_{UV}/m_{IR}=20$, and with $\gamma_{IR}$ and $\kappa$ tuned to $m_h = f_\pi/2$ and S=0.1\label{tab:nc4_nd2_l20}}
  	\end{table}

  	\begin{table}[!htbp]
  		\begin{tabular}{cc|cc|cccc|c|cc|cc}
  			\hline $N_c$ & $N_f$ & $\kappa$ & $\gamma_{IR}$  & $M_A$ GeV & $\tilde{g}$  & $\omega$ \\ \hline
  		 5 & 10. & 1.22 & 0.108 & 2994.82 & 7.40201 & 0.027 \\
  		 5 & 12. & 1.25 & 0.115 & 2987.77 & 7.41937 & 0.029 \\
  		 5 & 14. & 1.31 & 0.125 & 2962.05 & 7.4432 & 0.032 \\
  		 5 & 16. & 1.37 & 0.138 & 2949.85 & 7.4742 & 0.035 \\
  		 5 & 17.7 & 1.5 & 0.161 & 2911.46 & 7.521 & 0.041 \\
  		 5 & 19. & 1.6 & 0.181 & 2900.96 & 7.57054 & 0.046 \\
  		 5 & 20. & 1.78 & 0.216 & 2888.02 & 7.64511 & 0.053 \\
  		 5 & 20.5 & 2.15 & 0.282 & 2880.7 & 7.77428 & 0.064 \\
  		 5 & 20 & 3.1 & 0.432 & 2942.86 & 8.0188 & 0.081 \\
  		 5 & 19 & 4.5 & 0.63 & 3133.44 & 8.25156 & 0.094 \\
  		 5 & 17.7 & 6.9 & 0.965 & 3497.38 & 8.45762 & 0.119 \\
  			\hline
  		\end{tabular}
  		\caption{Data from the holographic model for $N_c=5$ one doublet $(N_D=1)$ with $\Lambda_{UV}/m_{IR}=20$, and with $\gamma_{IR}$ and $\kappa$ tuned to $m_h = f_\pi/2$ and S=0.1
  		\label{tab:nc5_nd1_l20}}
  	\end{table}

  \begin{table}[!htbp]
  	\begin{tabular}{cc|cc|cccc|c|cc|cc}
  		\hline $N_c$ & $N_f$ & $\kappa$ & $\gamma_{IR}$  & $M_A$ GeV & $\tilde{g}$  & $\omega$ \\ \hline
  	 3 & 2 & 2.12 & 0.168 & 2450.08 & 9.33511 & 0.131 \\
  	 3 & 4 & 2.23 & 0.18 & 2461.13 & 9.37285 & 0.130 \\
  	 3 & 9 & 2.4 & 0.192 & 2552.5 & 9.44938 & 0.120 \\
  	 3 & 10.6 & 2.54 & 0.2143 & 2642.26 & 9.52481 & 0.113 \\
  	 3 & 11 & 2.63 & 0.2245 & 2669.57 & 9.5557 & 0.111 \\
  	 3 & 12 & 3.1 & 0.279 & 2825.51 & 9.70417 & 0.099 \\
  	 3 & 12.1 & 3.22 & 0.294 & 2872.67 & 9.74188 & 0.097 \\
  	 3 & 12.1 & 4.15 & 0.396 & 3155.4 & 9.96588 & 0.082 \\
  	 3 & 12 & 4.48 & 0.4329 & 3257.74 & 10.0358 & 0.078 \\
         3 & 11.8 & 5.1 & 0.496 & 3416.16 & 10.1432 & 0.074 \\
  	 3 & 11.4 & 6.48 & 0.624 & 3701.95 & 10.3208 & 0.073 \\
  	 3 & 11 & 8.42 & 0.776 & 3988.24 & 10.4732 & 0.082 \\
  	 3 & 10.6 & 11.5 & 0.9694 & 4296.59 & 10.5911 & 0.105 \\
  		\hline
  	\end{tabular}\caption{Data from the holographic model for $N_c=3$ one doublet $(N_D=1)$ with $\Lambda_{UV}/m_{IR}=200$, and with $\gamma_{IR}$ and $\kappa$ tuned to $m_h = f_\pi/2$ and S=0.1\label{tab:nc3_nd1_l200}}
  \end{table}

  \clearpage

\end{document}